\documentclass[onecolumn,superscriptaddress,aps,preprintnumbers,longbibliography,amsmath,amssymb]{revtex4-2}

\usepackage{graphicx}
\usepackage{dcolumn}
\usepackage{bm}
\usepackage{physics} 
\usepackage{color,soul}
\begin{document}

\title{Brightening of dark trions in monolayer WS$_2$ via localization of surface plasmons}
\author{Sreyan Raha}
\thanks{Present Address : Universite de Toulouse, INSA-CNRS-UPS, LPCNO, 135 Avenue Rangueil, 31077, Toulouse, France}
\affiliation{Department of Physical Sciences, Bose Institute, EN 80, Sector V, Bidhannagar, Kolkata 700091, India.}

\author{Tara Shankar Bhattacharya}
\affiliation{Department of Physical Sciences, Bose Institute, EN 80, Sector V, Bidhannagar, Kolkata 700091, India.}
\affiliation{Kolaghat Government Polytechnic, Kolaghat 721134, India.}

\author{Indrani Bose}
\affiliation{Department of Physical Sciences, Bose Institute, EN 80, Sector V, Bidhannagar, Kolkata 700091, India.}


\author{Achintya Singha}
\email{Corresponding author: achintya@jcbose.ac.in}
\affiliation{Department of Physical Sciences, Bose Institute, EN 80, Sector V, Bidhannagar, Kolkata 700091, India.}

\begin{abstract}
Among all excitonic complexes in transition-metal dichalcogenides (TMDs), dark and semi-dark trions are poised to play a stellar role in future quantum technologies due to their long lifetimes, about two orders of magnitude greater than those of their bright counterparts. In monolayer (ML) tungsten disulphide (WS$_2$), accessing these states via a suitable brightening mechanism remains challenging, specially, at elevated temperatures. Here, we demonstrate the brightening of dark trions from ML WS$_2$ over the temperature range, 83 K–115 K, enabled by localized surface plasmon modes in a disordered gold substrate. The resulting photoluminescence (PL) spectrum reveals a distinct spectral doublet with the twin peaks of semi-dark and bright trion states, separated by $\sim$ 45 meV. The origin of the semi-dark trion state lies in intervalley electron-electron scatterings, while its visibility in the PL spectrum is made possible by the enhanced out-of-plane electromagnetic field associated with plasmon localization. We also report on the negative degree of circular polarization in ML WS$_2$ at the energy of the semi-dark trion state. Our results establish a scalable plasmonic route to access valley-polarized semi-dark trions, opening new opportunities for quantum and valleytronic applications.

\end{abstract}

\flushbottom
\maketitle
\newpage
\section{Introduction}
Monolayer (ML) transition metal dichalcogenides (TMDs) represent a distinctive class of atomically thin semiconductors known for their remarkable exciton dynamics and spin-valley coupled properties. The characteristic features of the ML materials allow for a number of unique physical properties \cite{schaibley2016valleytronics,Xu2014}. A direct band gap separates the lowest conduction band (CB) and the highest valence band (VB) states at the $\pm$K valleys of the electronic band structure. Coulomb interaction along with reduced dielectric screening facilitate the formation of an optically generated exciton, a bound pair of an electron and a hole, across the gap. A combination of broken inversion symmetry and time-reversal symmetry (degenerate energy states with opposite spin orientations) of the band states at the $\pm$K valleys give rise to observable valley-specific phenomena, e.g., valley-selective optical excitation with circularly polarized light \cite{schaibley2016valleytronics,Xu2014}. Additionally, the presence of strong spin-orbit coupling splits the VB and CB states with the splitting magnitude much larger in the case of the VB. The spin splitting of the band states couples the spin and valley degrees of freedom and  allows for the formation of excitonic complexes like dark excitons and dark/bright  trions. These attractive features, taken together, are central to emerging  valleytronic applications in optoelectronics, spintronics and quantum information-based technologies.

While bright excitons in ML TMDs have been extensively investigated, increasing attention is now devoted to their dark counterparts, particularly in tungsten-based materials such as WS$_2$ and WSe$_2$ \cite{Zhang2017,Park2018,Gelly2022,PhysRevB.111.155409,zhou2017probing}. In these systems, the lowest-energy exciton turns out to be an optically inactive dark exciton with the spins of the associated CB and VB states being antiparallel. The dark state is populated via the thermal relaxation of an optically generated bright exciton to the lower energy state. The negatively charged trions in ML TMDs, which are bound complexes of two electrons and a hole, are of three types: bright intravalley or singlet trion (X$_{\mathrm{S}}^{-}$), intervalley or triplet trion (X$_{\mathrm{T}}^{-}$) and the dark trion (X$_{\mathrm{D}}^{-}$), as depicted in Figs. 1(a–c). Owing to its charged three-particle structure, a trion encodes more information than a single  electron with the added advantage that some of its properties  are electrically tunable. The lifetime of an excitonic complex is finite due to both radiative and non-radiative decay processes.  Since a dark trion, being optically inactive, does not undergo radiative decay, its lifetime is about two orders of magnitude larger than that of a bright trion, making it a promising candidate for optoelectronic and quantum information applications \cite{Robert2021,PhysRevLett.123.027401}.  

\begin{figure}[h]
\centering
 \includegraphics[height=10cm]{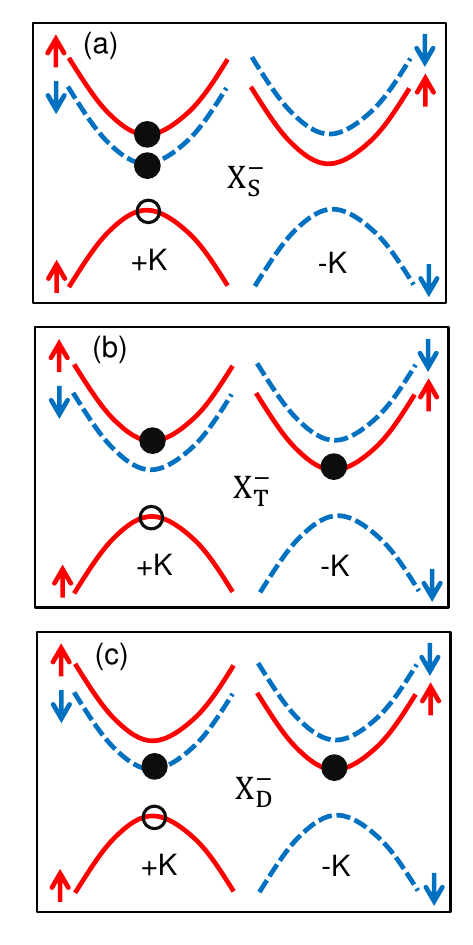}
  \caption{(a) singlet bright trion (X$_{\mathrm{S}}^{-}$), (b) triplet bright trion (X$_{\mathrm{T}}^{-}$), (c) dark trion (X$_{\mathrm{D}}^{-}$).}
\end{figure}

\begin{figure}
\centering
 \includegraphics[height=8cm]{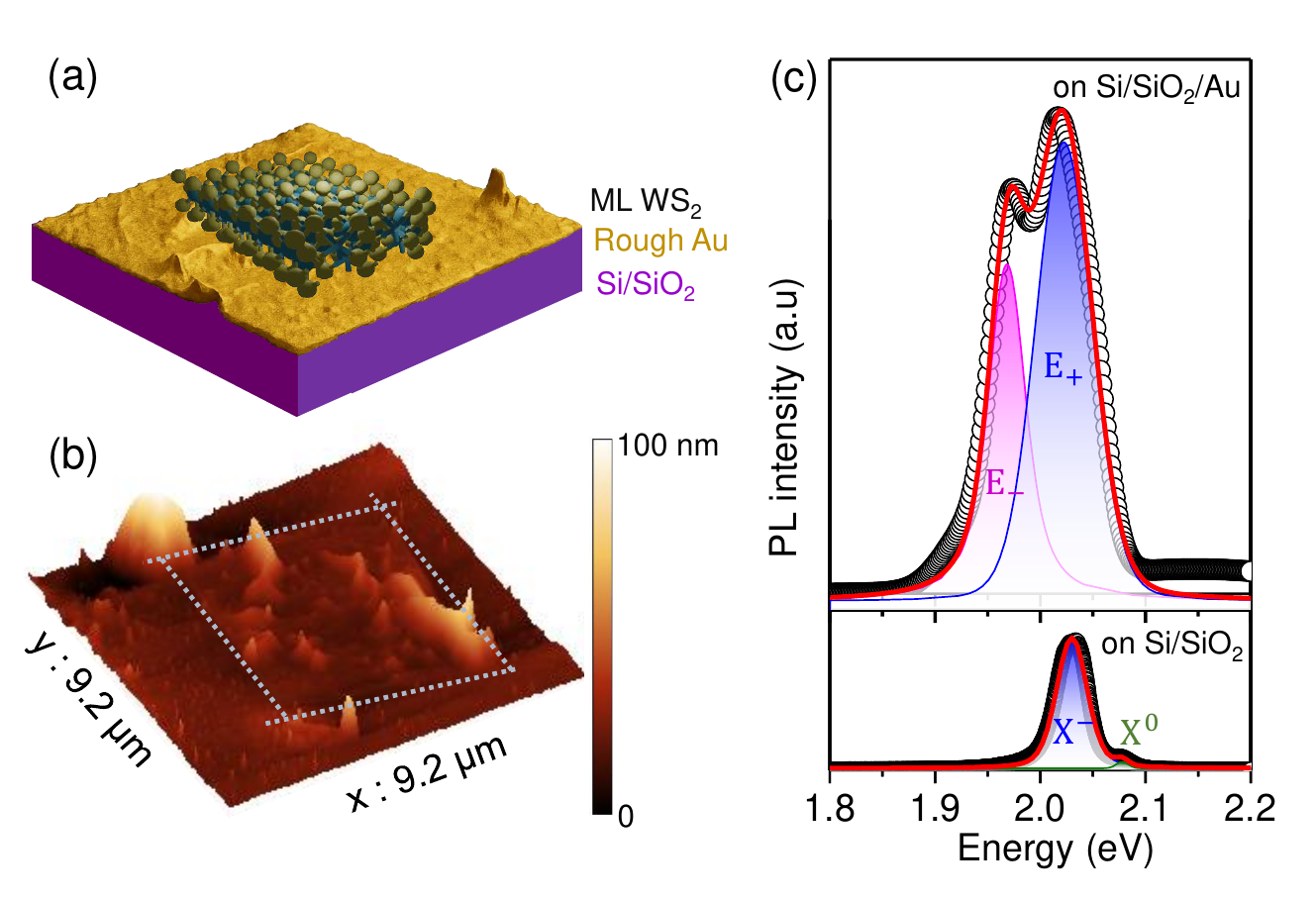}
  \caption{(a) A schematic diagram of the experimental arrangement of the ML WS$_2$ sample on a disordered Au film with an underlying  Si/SiO$_2$ substrate. (b) A 3D AFM profile of the rough gold surface with the sample covering the dotted rectangular area. (c) PL spectra at T= 83 K of ML WS$_2$ placed on the Si/SiO$_2$ substrate, and on the Si/SiO$_2$/Au substrate, as labeled in the figure.}
\end{figure}

Such applications, however, require at least a partial optical activation (“brightening”) of the dark excitonic complexes for their detection. Various approaches have been explored so far to achieve the brightening, e.g, by using magnetic fields, through strain engineering and via coupling to nanostructures with enhanced light-matter interactions \cite{Zhang2017,Park2018,Gelly2022,Robert2021,PhysRevLett.129.027402,PhysRevLett.131.116901,PhysRevLett.115.257403}. Though a number of studies have been carried out on the brightening of dark excitons and  trions in  ML WSe$_2$ over an extended temperature range,  the situation is markedly different for ML WS$_2$ \cite{PhysRevLett.115.257403}.

In this paper, we report on the optical activation of the dark trion state in ML WS$_2$ at elevated temperatures (83 K-115 K) by exploiting Anderson-like localization of surface plasmons in a disordered Au film in contact with the TMD ML \cite{10.1063/5.0001451,PhysRevLett.82.4520}. Through temperature-dependent photoluminescence (PL) spectroscopy, we identify a distinct spectral doublet in the temperature range of 83 K–115 K, with the lower and higher energy peaks attributed to the semi-dark ($\lvert{E_-}\rangle$) and bright trions ($\lvert{E_+}\rangle$), respectively. A theoretical model \cite{danovich2017dark} suggests the semi-dark trion state ($\lvert{E_-}\rangle$) to be a mixed state of the dark ($\lvert{D}\rangle$) and bright trion ($\lvert{B}\rangle$) states coupled via electron-electron (e-e) intervalley scatterings. The Anderson-like localized surface plasmons in the disordered Au film contribute to the optical brightening of the dark trions and also in the enhancement of the PL signals from bright trions. Using polarization-resolved PL spectroscopy, we further show that the degree of circular polarization ($P_{\mathrm{c}}$) is negative at the peak energy of the semi-dark trion state ($\lvert{E_-}\rangle$).

\section{Results and Discussions}
Fig. 2(a) schematically shows the WS$_2$ ML deposited on a disordered Au film supported by a Si/SiO$_2$ substrate. See Section I of supplementary information for the details of the sample preparation. The incident laser excites collective oscillations of free electrons in the Au surface generating surface plasmons. The structural disorder in the Au film gives rise to Anderson-like localization of these modes in the form of localized surface plasmons (LSPs) \cite{10.1063/5.0001451,PhysRevLett.82.4520}. Fig. 2(b) exhibits a 3D AFM image of the rough gold surface, with the region enclosed by the white dotted line indicating the area covered by the WS$_2$ ML.

Fig. 2(c) compares the PL spectra at 83 K of the ML WS$_2$  sample on Si/SiO$_2$, with and without the intervening rough Au film. The PL emission spectrum of ML WS$_2$ on the Si/SiO$_2$ substrate has contributions from both bright excitons (X$^0$) and negatively-charged bright trions (X$^-$) (also see Fig. S1(a)). The peak energy difference between the bright exciton and bright trion is found to be $\sim$ 40 meV, in the range of quoted values in literature \cite{jadczak2017probing,jadczak2019room,kesarwani2022control}.

Surface plasmon-induced hot electron injection into ML TMD materials has been extensively investigated in recent times \cite{tugchin2023photoluminescence,wen2022pathways}. The studies further establish that the injected
electrons interact with the existing neutral excitons and convert them into negatively charged trions. Various optical control strategies for exciton-trion conversion have also been explored \cite{lee2023all,jeong2025tip}. A
schematic diagram illustrating the hot electron injection process is given in Fig. S2. In the light of the reported experimental evidence, we suggest that the presence of the Au substrate in our experiment promotes n-doping of the WS$_2$ sample, via hot electron injection, resulting in the neutral exciton population becoming negligible as the temperature is lowered, with most of the excitons effectively converted into trions (Fig. 2(c)). This allows one to exclusively focus on the dynamics of the trions. In the presence of the Au film, a notable PL enhancement is observed, consistent with plasmon-assisted emission \cite{wen2022pathways,lee2023all,jeong2025tip, PhysRevB.100.235438,chumkiPRB,Rahatuning}. The PL emission spectrum is a doublet with a peak separation of about 45 meV.

To exclude localized strain as the origin of the new spectral feature, we performed spatially resolved PL measurements across different regions of the sample. As shown in Fig. 3(a), the spectral shape remains uniform across the entire sample along the line 1–5. We have also  carried out position-dependent Raman measurements, described in Section IV of the Supplementary Information, and probed the local topography using AFM (Section V, Supplementary Information) to confirm that the observed optical response of the sample is not strain-induced.

We further examined the nature of the two peaks through laser power-dependent PL measurements at 83 K with the spectra shown in Fig. S5. The log-log plot of the integrated PL intensity, I versus the laser power, P (Fig. 3(b)) reveals a linear relationship, I $\propto$ P$^k$, with k $\sim$ 1.2 ($\sim$ 1) for the higher (lower) energy peak, ruling out emissions due to defect-bound recombination processes which typically show sub-linear scaling (k $<$ 1) \cite{PhysRevLett.121.057403}.

In ML TMDs, bright excitons and trions have in-plane (IP) dipole moments, while dark excitons and trions carry out-of-plane (OUP) dipole moments. In standard optical set-ups for probing circularly polarized PL, light is incident along a direction normal to the sample so that the associated electric field is IP. As per Fermi’s Golden Rule (FGR), the radiative transition, i.e., the PL emission rate is proportional to $|{\bf p}\cdot{\bf E}|^{2}$ where \textbf{p} is the electric dipole moment of the emitter (exciton/trion) and \textbf{E} the optical electric field vector so that only the bright excitons and trions are optically active. As discussed by Purcell \cite{Park2018}, the spontaneous emission rate of a quantum emitter is considerably enhanced in a confined electromagnetic environment provided by nanostructures. Such an environment provides a large density of  available  electromagnetic states for the quantum emitter to emit into, raising the emission rate considerably. In the light of the above discussion, we now  elucidate the role of the rough Au film in the generation of an enhanced PL signal.  
The LSPs generated in the rough Au film have a prominent OUP electric field pointing in the z-direction. This is demonstrated in Fig. S6(b), through numerical simulations of the electric field intensity, ($|{E_z}|^{2}$), associated with the LSPs. The figure displays localized hot spots of high field intensity, above the experimentally measured AFM topography, with an enhancement factor in the range of 100-200. More discussion on the role of the rough Au film can be found in Section VIII of the Supplementary Material. The OUP electric field couples to the OUP dipole moment of the dark trion enabling a radiative transition (the factor $|{\bf p}\cdot{\bf E}|^{2}$ in FGR) and hence the “brightening” of the dark trion. The Anderson-type localization of the surface plasmons in the Au film gives rise to enhanced near-fields as well as an increased local density of states (LDOS) (the Purcell effect) giving rise to a marked enhancement of the PL signal. Furthermore, the less dominant IP components of the plasmonic field couple to the IP electric dipole moments of the bright trions so that the PL emission from the bright trions is also enhanced \cite{mueller2023photoluminescence}. The brightening of the negatively-charged dark trion with detectable intensity necessitates the doping of the sample with electrons so that the process of trion formation becomes favorable \cite{RAHA2025109373}. A doped electron lowers its energy by binding to an exciton and forming thereby a trion. In our experiment, the doping is
predominantly mediated via the Au film, which also provides the plasmonic
environment for enhanced PL signals, so that in its absence the PL signal from
the bright trion is much diminished in intensity (Fig. 2(c)).

The two PL peak energies of the spectral doublet (Fig. 2(c)), denoted by $E_{-}$ and $E_{+}$ with $E_{-}<E_{+}$, correspond to mixed states of dark ($\lvert{D}\rangle$) and bright ($\lvert{B}\rangle$) trions. The coupled state with energy $E_-$ ($E_+$) has the dark (bright) trion as the dominant component and can be characterized as a semi-dark ($\lvert{E_-}\rangle$) (bright ($\lvert{E_+}\rangle$)) trion state. To investigate the temperature dependence of the peak intensities, PL measurements were performed from 83 K to 293 K. The false color map of normalized PL intensity in Fig. 3(c) shows the emergence of $E_{-}$ near 115 K, while Fig. 3(d) shows the integrated intensities of both the peaks as a function of temperature. The semi-dark trion ($\lvert{E_-}\rangle$), being the lower energy state, does not require thermal activation and its PL intensity is expected to be higher than that of the bright trion at low temperatures. On the other hand, the bright states are dominant at elevated temperatures due to thermal activation from the dark states. This thermal behavior can be modeled in the framework of a two-level system by the following equation based on the Boltzmann distribution:

\begin{figure}
\centering
 \includegraphics[height=10cm]{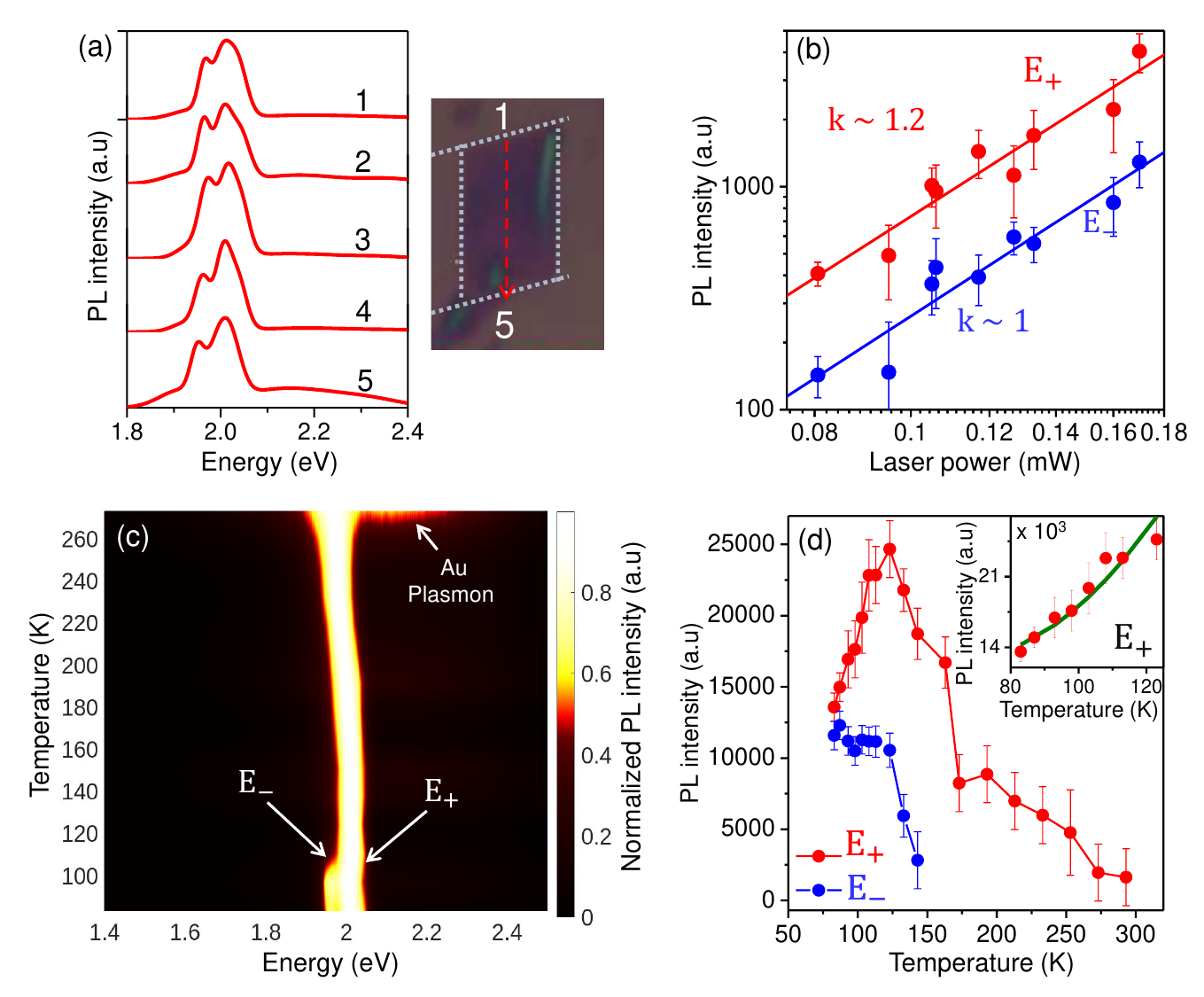}
  \caption{(a) PL spectra at 83 K from different regions of the sample, along the line  shown in the right panel. (b) PL integrated intensity of $E_+$ and $E_-$peaks at 83 K as a function of laser power. (c) False color plot for normalized PL intensity as a function of temperature from ML WS$_2$ on rough Si/SiO$_2$/Au substrate. (d) Integrated PL intensity of $E_+$ and $E_-$ states as a function of temperature.}
\end{figure}

\begin{equation}
I(T) = A \frac{e^{(-\frac{\Delta E}{k_BT})}}{1 + e^{(-\frac{\Delta E}{k_BT})}} + B,
\end{equation}

where $I(T)$ is the PL intensity of the $E_{+}$ peak, $A$ is a proportionality constant, $B$ is a temperature-independent background term, $k_{B}$ is the Boltzmann constant, and $\Delta E$ the energy separation between the two states. The functional form shown in equation (1) is similar to that proposed in \cite{arora2020dark} to describe the evolution of the trion PL intensity as a function of temperature for the two W-based
materials.
As shown in the inset of Fig. 3(d), the model provides a reasonable fit to the experimental data in the 83–115 K range. The extracted value of $\Delta$E $\sim$ 42 ($\pm$ 5) meV is consistent with the experimentally observed energy separation between the two components of the doublet. The ascending part of the PL intensity versus temperature plot indicates thermal activation from the semidark trion state $\lvert{E_-}\rangle$ to the higher energy bright trion $\lvert{E_+}\rangle$ followed by a descending
part reflecting the thermal breakup of the trions at elevated temperatures.
At T $\sim$ 113 K, the PL intensity of the $E_{+}$ peak is enhanced by a factor of $\sim$5 from the value obtained in the case of ML WS$_2$ on a Si/SiO$_2$ substrate. In the absence of the Au component of the substrate, the bright trion peak appears in the PL spectrum with much diminished intensity (Fig. 2(c)). The PL intensity of this bright trion decreases with increasing
temperature (Fig. S10).

As discussed previously, it is likely that the observed doublet in the PL emission   spectrum originates from the mixing of the dark ($\lvert{D}\rangle$) and the bright ($\lvert{B}\rangle$) trion states. The radiative recombination of the dark trion is generally brought about through its partial conversion to an optically allowed emission mode, characterized as the semi-dark trion. The theoretical model of Danovich et al. \cite{danovich2017dark} suggests a possible mechanism to achieve this, via intervalley electron-electron (e-e) scatterings which mix the states $\lvert{B}\rangle$ and $\lvert{D}\rangle$.  Fig. 4(a) shows a schematic diagram of the intervalley e-e scattering in which the two electrons occupying the lower CB subbands at the $\pm$K valleys, constituting the dark trion state $\lvert{D}\rangle$, are scattered to the upper CB subbands of the opposite $\mp$K valleys, resulting in the formation of an excited trion state configuration, represented by the bright triplet trion state $\lvert{B}\rangle$. The mixed state $\lvert{E_-}\rangle$, a linear combination of the $\lvert{D}\rangle$ and $\lvert{B}\rangle$ states, has the dark (bright) trion state as the dominant (minority) component and hence designated as the semi-dark trion state. This state derives a small but non-zero oscillator strength from the bright trion component so that radiative PL emission, via the recombination of the e-h pair in the +K valley, is possible. Similarly, the mixed state $\lvert{E_+}\rangle$ has the bright (dark) trion state as the dominant (minority) component.

There is by now considerable experimental evidence for the formation of the semi-dark trion state via intervalley e-e scattering \cite{doi:10.1021/acs.nanolett.0c05021,chand2023interaction,tu2019experimental} and other related mechanisms \cite{PhysRevB.111.155409}. The PL spectra were obtained at low temperatures, e.g., at 5 K \cite{doi:10.1021/acs.nanolett.0c05021} and 7 K \cite{chand2023interaction} for the ML WS$_2$ sample and at 4 K \cite{PhysRevB.111.155409} and in the range of temperatures, 12 K – 60 K, for the ML WSe$_2$ sample \cite{tu2019experimental}. In each case, the semi-dark trion has a distinct presence in the PL spectrum. In some of the experiments \cite{PhysRevB.111.155409,doi:10.1021/acs.nanolett.0c05021,chand2023interaction} and in the presence of a magnetic field, the dark trion peak, made visible by the field, is seen in the PL spectrum shifted by an energy of $\sim$12 meV above the peak energy of the semi-dark trion state, a unique feature of the PL spectra of both the W-based materials at low temperatures. 
The energy shift can be understood by referring to Fig. 4(a). For radiative recombinations in semiconductor crystals, the recombination process has to satisfy energy, momentum and spin conservation rules. The coulomb-induced intervalley e-e scatterings shown in Fig.4(a) are spin-conserving. Radiative recombination of the e-h pair takes place in the +K valley with the upconversion of the extra (spectator) electron from the lower CB subband of the +K valley to the upper CB subband of the –K valley to conserve momentum. The energy of this extra electron is above that of the dark trion energy E$_{D^-}$ by the amount   $\Delta_{SO}$, the CB spin-orbit splitting energy. Due to the requirement of energy conservation, the PL emission energy, $E_{-}$, of the semi-dark trion  is given by E$_-$ = E$_{D^-}$ - $\Delta_{SO}$. Recent single-particle estimates, extracted from magneto-optical spectroscopy and transport measurements yield $\Delta_{SO}$=12$\pm$1 meV for WS$_2$ and $\Delta_{SO}$=12$\pm$0.5 meV for WSe$_2$ \cite{ren2023measurement,kapuscinski2021rydberg}, the values notably smaller than those from band structure calculations (e.g., $\Delta_{SO}$=32 meV for WS$_2$ as quoted in \cite{danovich2017dark}). The experimental estimate of $\Delta_{SO}$=12 meV is in close agreement with the reported peak energy differences of the brightened dark and the semi-dark trions in the PL spectra of both ML WS$_2$ and  ML WSe$_2$ \cite{PhysRevB.111.155409,doi:10.1021/acs.nanolett.0c05021,chand2023interaction}, conferring validity on the model based on intervalley e-e scattering. In these experiments carried out at temperatures in the range of 4K -7K, four trion peaks could be observed in the PL spectrum associated successively with the semi-dark (lowest energy), dark (Fig. 1(c)), bright singlet and triplet (Figs. 1(a) and (b)) trions. Our PL measurements on ML WS$_2$ were carried out at elevated temperatures and could not resolve the four distinct trion peaks due to thermal effects. The PL spectrum exhibits only a doublet with two prominent peaks (Fig. 2(c)) with the peak energies $E_-$ and $E_+$ associated with the semi-dark and bright trion states respectively. A four-component analysis of the PL doublet at 83 K, reported in Section XI of Supporting Information, shows overlapping PL signals from all the four trions (Fig. S11)  with the peak energy differences close to the experimental estimates.

In the low-temperature experiments \cite{PhysRevB.111.155409,doi:10.1021/acs.nanolett.0c05021,chand2023interaction,tu2019experimental}, the visibility of the  semi-dark trion peak is considerably weakened as the temperature is raised due to the thermal activation  of the semi-dark trion to the higher energy bright trion. For example, in the case of ML WSe$_2$, the semi-dark trion peak loses visibility around 60 K. The inclusion of the rough Au film in our substrate has the advantage of extending the temperature range for the detection of the semi-dark trion, via the LSP-mediated amplification of the weak signal. The semi-dark trion state has the dark (bright) trion as the dominant (minority) component and, in the absence of the Au film, yields a PL emission signal solely from the bright trion component. In the presence of the film, the OUP electric field of the LSPs couple to the OUP electric dipole moment of the dark trion making PL emission from the dark component possible. Furthermore, the less dominant IP component couples to the IP electric dipole moment of the bright trion component enhancing its PL emission signal. The new contributions amplify considerably the net PL signals from both the semi-dark and bright trion states, $\lvert{E_-}\rangle$ and $\lvert{E_+}\rangle$ (Fig. 2(c)). In the earlier experiments, the brightening of the dark component was achieved using a magnetic field \cite{doi:10.1021/acs.nanolett.0c05021}. As proposed by Danovich et al. \cite{danovich2017dark}, the bright ($\lvert{B}\rangle$) and dark trion ($\lvert{D}\rangle$) states in a monolayer sample are coupled through intervalley e-e scatterings giving rise to two mixed states, a semi-dark trion state ($\lvert{E_-}\rangle$) and a state similar to the bright trion state ($\lvert{E_+}\rangle$). The coupling can be described by a 2$\times$2 matrix

\begin{equation}
H = 
\begin{pmatrix}
E_b & \mu \\
\mu^* & E_d
\end{pmatrix},
\end{equation}
in the basis states $\lvert B \rangle$ and $\lvert D \rangle$ representing the bright and dark trion states with respective energies $E_b$ and $E_d$ with $\mu$ being the coupling parameter.
The energies $E_\pm$  of the higher and lower peak positions (Fig. 2(c)) are given by
\[
E_{\pm} = \frac{E_b + E_d}{2} \pm \frac{1}{2} \sqrt{(E_b - E_d)^2 + 4|\mu|^2}
\]  
with the corresponding state vectors given by $\lvert{E_+}\rangle$ and $\lvert{E_-}\rangle$ representing the semi-dark and bright trion mixed states, respectively. The energy gap between the two peaks is given by
\[
\Delta_D = E_{+} - E_{-} = 
\sqrt{(E_b - E_d)^2 + 4|\mu|^2}
= \sqrt{(2\Delta_{SO})^2 + 4|\mu|^2}
\]

where $E_b - E_d \approx 2\Delta_{SO}$ \cite{danovich2017dark}. With the CB spin-orbit splitting energy $\Delta_{SO}$=12 meV and the coupling parameter $\mu$=18 meV (from \cite{danovich2017dark}), one obtains $\Delta_{D}$=43.3 meV, in close agreement with the experimentally observed value of  $\Delta_{D}$=45 meV. Taking $\Delta_{SO}$=13 meV (within the experimental certainty) gives $\Delta_{D}$=44.4 meV. The quantitative analysis firmly links the origin of the observed PL spectral doublet (Fig. 2(c)) to intervalley e-e scattering, with the extracted peak energy splitting in good agreement with the theoretical estimate.  A similar agreement between theoretical and experimental estimates of $\Delta_D$ has been demonstrated in the case of ML WSe$_2$ \cite{tu2019experimental}. At this point, we emphasize that while  intervalley e-e scatterings govern the formation of the mixed trion states, ($\lvert{E_-}\rangle$) and  ($\lvert{E_+}\rangle$), the PL emission intensity depends on the coupling between the electric dipole  moment of the emitter and the plasmonic near-field, the electromagnetic environment (Purcell effect) as well as the temperature.

In the low temperature experiments, the energy difference $\Delta_{D}$ between the bright triplet and semi-dark trion peaks has the approximate values of 39 meV (WSe$_2$) \cite{PhysRevB.111.155409}, 37 meV (WS$_2$) \cite{doi:10.1021/acs.nanolett.0c05021}, 38 meV  (WS$_2$) \cite{chand2023interaction} and 55 meV (WSe$_2$) \cite{tu2019experimental}. In the first three references, hBN encapsulated samples were used and in the last case the sample substrate used was Si/SiO$_2$. In our experiment, the sample ML WS$_2$ was deposited on the rough Au/Si/
SiO$_2$ substrate. The estimates of $\Delta_{D}$ from experimental measurements are: 45
meV (from PL spectrum) and 42($\pm$5)meV from fitting the thermal activation formula (Eq. (1)) to the ascending part of the plot of the bright
trion PL emission intensity versus temperature (Fig. 3(d)).
In all the cases, mechanisms based on intervalley e-e scatterings were proposed to explain the PL spectral features.

\begin{figure}
\centering
 \includegraphics[height=12cm]{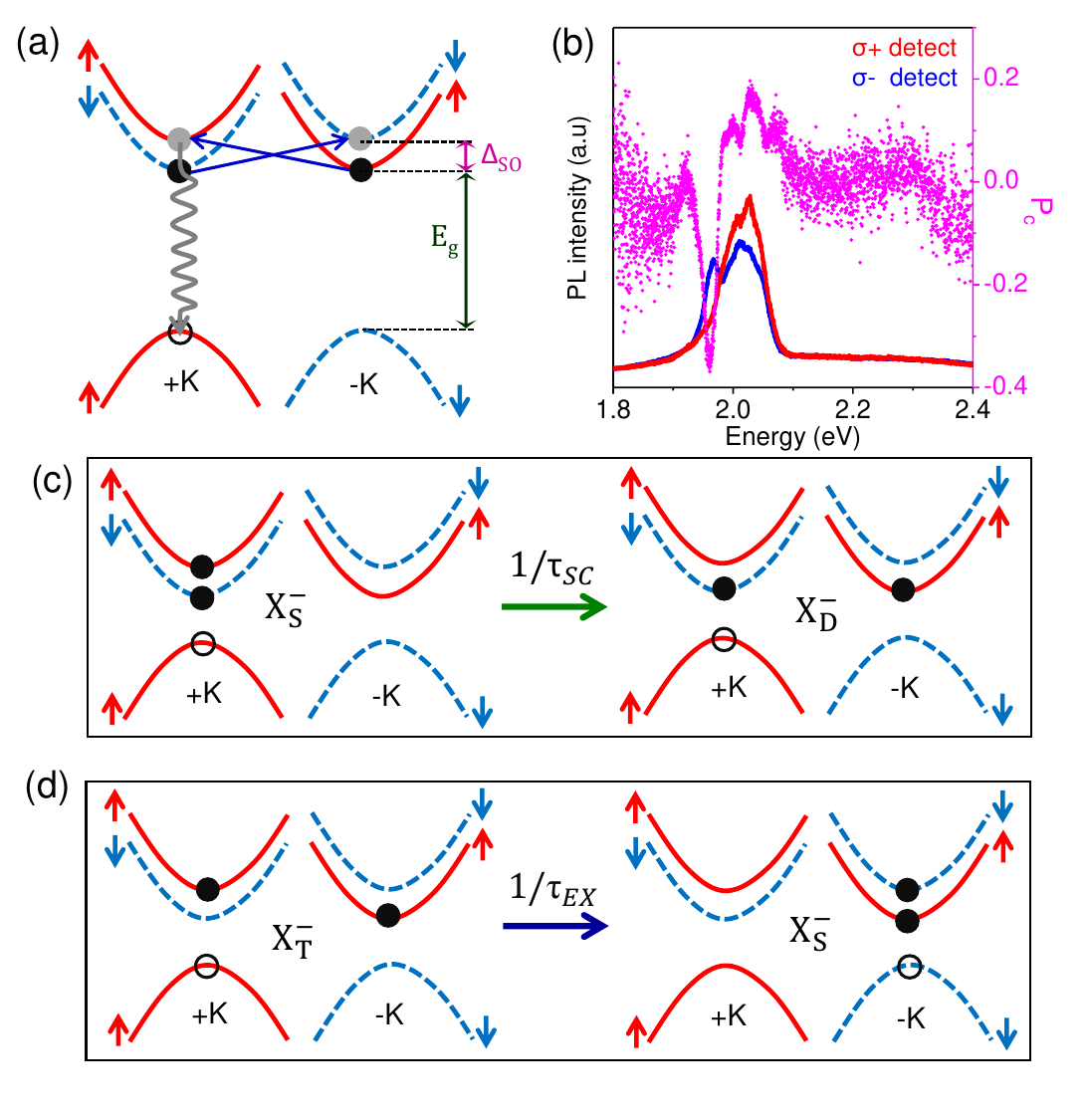}
  \caption{Schematic diagram of the intervalley scattering processes that couple the dark and bright states of trions. $E_{g}$ is the band gap and $\Delta_{SO}$ stands for the CB spin-splitting. (b) Circular polarization-resolved PL spectrum of ML WS$_2$ on disordered Si/SiO$_2$/Au substrate, as detected in the $\sigma+$ and $\sigma-$ channels on excitation  by $\sigma+$ 488 nm (2.54 eV) laser irradiation. (c) Spin conserving scattering process from X$_{\mathrm{S}}^{-}$ to X$_{\mathrm{D}}^{-}$. (d) Scattering from X$_{\mathrm{T}}^{-}$ to X$_{\mathrm{S}}^{-}$ via e-h exchange interaction.}
\end{figure}

To probe the valley dynamics of the trion states, we performed circular polarization-resolved PL measurements. The sample was excited with $\sigma+$ light, and the PL signals were detected in the $\sigma+$ and $\sigma-$ detection channels. The degree of circular polarization, $P_{\mathrm{c}}$, is calculated as

\begin{equation}
P_{\mathrm{c}} = \frac{I(\sigma+) - I(\sigma-)}{I(\sigma+) + I(\sigma-)},
\end{equation}

where $I(\sigma+)$ and $I(\sigma-)$ correspond to the intensity of the PL signal detected in the $\sigma+$ and $\sigma-$ channels, respectively. The $P_{\mathrm{c}}$ (magenta curve) as a function of energy, shown in Fig.~4(b), reveals a negative $P_{\mathrm{c}}$ value ($-35\%$) around 1.96~eV, corresponding to the peak energy $E_{-}$ of the semi-dark trion state $\lvert{E_-}\rangle$, and a positive $P_{\mathrm{c}}$ ($18\%$) around the position of $E_{+}$, associated with the bright trion state $\lvert{E_+}\rangle$. 

This result is similar to that reported by Zhang et al. \cite{Zhang2017}, on magnetic-field brightening of the dark (gray) exciton and trion in ML WSe$_2$. In their case, the brightened dark excitonic complexes exhibited negative $P_{\mathrm{c}}$, i.e., $I(\sigma-) > I(\sigma+)$, in contrast to  the case of  bright states, where the PL emission is co-polarized with the laser excitation. Although the precise mechanism underlying the observation could not be specified, the relaxation of bright excitons to dark excitons through intervalley electron scattering was speculated to be a possible origin.

To  explain the origin of the negative valley polarization, one needs to take  into account  the singlet and triplet bright trions as well as the dark trions. As suggested in a theoretical study by Fu et al. \cite{FuAPL2019}, the reversal of the valley polarization (VP) could arise from the conversions between the dark and bright trion states. The reversal of the VP studied by Fu et al. \cite{FuAPL2019} refers to the polarization of the singlet trion X$_{\mathrm{S}}^{-}$ (negative $P_{\mathrm{c}}$) in contrast to that of the triplet trion X$_{\mathrm{T}}^{-}$ (positive $P_{\mathrm{c}}$). The splitting of a bright trion peak in the PL spectrum into a pair of singlet (lower energy) and triplet trion peaks has been demonstrated experimentally for both WSe$_2$ and WS$_2$ MLs, with the magnitude of the splitting measured to be approximately 6--7~meV \cite{Plechinger2016}. Experimental evidence for the negative $P_{\mathrm{c}}$ of the singlet trion X$_{\mathrm{S}}^{-}$ has been reported only for WSe$_2$ MLs \cite{Robert2021}.

Two dynamical processes play a dominant role in  the interconversions between the trion states, as illustrated in Figs. 4(c,d). In the first process (Fig. 4(c)), X$_{\mathrm{S}}^{-}$ scatters into the dark trion X$_{\mathrm{D}}^{-}$ via spin-conserving transfer (rate given by 1/$\tau_{SC}$) of an electron from the upper CB subband in the +K valley to the lower CB subband in the –K valley \cite{PhysRevLett.117.257402}. An explicit value of $\tau_{SC}$ has not been reported so far but, as pointed out in Refs \cite{FuAPL2019,PhysRevLett.117.257402,mccormick2017imaging,PhysRevB.95.235408}, the transition from X$_{\mathrm{S}}^{-}$ to X$_{\mathrm{D}}^{-}$ turns out to be  the most favorable process in the inter-conversions between the bright and dark trions \cite{FuAPL2019,PhysRevLett.117.257402,mccormick2017imaging,PhysRevB.95.235408}. This implies that the scattering rate 1/$\tau_{SC}$ is much larger than the singlet trion recombination rate. In the second process (Fig. 4(d)), the excitonic e-h pair part of  X$_{\mathrm{T}}^{-}$ in the +K valley is transferred  (rate given by 1/$\tau_{EX}$, $\tau_{EX}$ $\sim$ 4 ps at 13 K) to the –K valley, forming X$_{\mathrm{S}}^{-}$ \cite{PhysRevLett.117.257402}. This process is mediated through the  e-h exchange coupling, which is known to provide an efficient channel for valley depolarization \cite{FuAPL2019,PhysRevLett.117.257402,mccormick2017imaging,Vaclavkova_2018,lyons2019valley}. The other possible dynamical processes require spin flips, thermal activation or simultaneous transfers of three particles, which are considerably less efficient in bringing about valley depolarization \cite{FuAPL2019,PhysRevLett.117.257402,mccormick2017imaging,PhysRevB.95.235408,Vaclavkova_2018,lyons2019valley}.

To understand the valley population imbalance, we re-express the $P_{\mathrm{c}}$ as

\begin{equation}
P_{\mathrm{c}} = \frac{N(+K) - N(-K)}{N(+K) + N(-K)},
\end{equation}

where N(+K) and N(-K) denote the numbers of singlet trions in the +K and -K valleys respectively. The dominant (minority) component of the semi-dark trion state {$\lvert{E_-}\rangle$} is the dark (bright) trion state. In this state, there is a spectral overlap of the bright trion state, and hence its singlet component (X$_{\mathrm{S}}^{-}$), with the dark trion state (X$_{\mathrm{D}}^{-}$). The dominance of the dark component implies a significant conversion of X$_{\mathrm{S}}^{-}$ to X$_{\mathrm{D}}^{-}$ in the +K valley. Moreover, the e-h exchange coupling facilitates the conversion of X$_{\mathrm{T}}^{-}$ in the +K valley to X$_{\mathrm{S}}^{-}$ in the -K valley. Thus, the occupation number of X$_{\mathrm{S}}^{-}$ in the -K valley exceeds that in the +K valley, i.e., $N(-K) > N(+K)$. Given that the optical transitions from  the -K valley yield $\sigma-$ polarized emission, the PL signal is characterized by a negative value of $P_c$. In the bright trion state $\lvert{E_+}\rangle$, the dominant component comes from the bright trion itself while the dark trion is a minority constituent. Since the coupled state is predominantly bright, i.e., the proportion of dark states is small, the population transfer from X$_{\mathrm{S}}^{-}$ to X$_{\mathrm{D}}^{-}$ is suppressed, and the depletion of X$_{\mathrm{S}}^{-}$  from the +K valley is not significant. Furthermore, the $P_{\mathrm{c}}$ of the bright trion component is positive resulting in a net positive value of $P_c$.

\section{Conclusion}
In conclusion, our study demonstrates the brightening of dark trions and the amplification of the PL emission intensity, over the temperature range of 83 K -115 K, in a WS$_2$ ML sample. The earlier experiments \cite{doi:10.1021/acs.nanolett.0c05021,chand2023interaction} on ML WS$_2$, carried out at considerably lower temperatures (5 K and 7 K), reported a distinct semi-dark trion (partially brightened dark trion) emission line in the PL spectrum, the origin of which lies in intervalley e-e scatterings (Fig. 4(a)). The key prediction of the scattering mechanism, that the peak energy difference between the dark ($\lvert{D}\rangle$) and semi-dark ($\lvert{E_-}\rangle$) trions is of the order of, $\Delta_{SO}$, the CB spin-orbit splitting, has been experimentally confirmed \cite{PhysRevB.111.155409,doi:10.1021/acs.nanolett.0c05021,chand2023interaction}. The four-component analysis of our PL doublet also reproduces the result. As reported in \cite{tu2019experimental}, the PL emission line from the semi-dark trion becomes weaker, losing visibility thereby, at elevated temperatures due to the thermal activation of the lowest-energy semi-dark trion to the higher-energy bright trion. The inclusion of the rough Au film in our experiment serves a dual purpose: to n-dope the WS$_2$ sample, via hot electron transfer, promoting trion formation, an alternative to n doping through electrostatic gating \cite{PhysRevB.111.155409} and also to amplify the PL signals from both the semi-dark and bright trion states, $E_-$ and $E_+$. To summarize, our experimental study illustrates how a combination of intervalley e-e scattering and plasmonic coupling enables the detection of the PL signal emissions from the different species of trions in ML WS$_2$, at elevated temperatures. The quantitative estimates emerging from our study are consistent with experimental results. The experimental observation of a negative $P_c$ around the PL emission energy $E_-$ is explained on the basis of available experimental evidence on trion dynamics in tungsten-based TMD  materials. Beyond WS$_2$, our approach establishes a scalable strategy for optical access to dark excitonic complexes in two-dimensional semiconductors, giving rise to new possibilities for valleytronic and nanophotonic applications.

\setcounter{figure}{0}

\renewcommand{\thefigure}{S\arabic{figure}}

\newpage
\hrule
\begin{center}
\textbf{\Large Supporting Information}
\vspace{0.1 in}
\hrule
\end{center}


\begin{center}
\textbf{Section I : Methods}
\end{center}

ML WS$_{2}$ was mechanically exfoliated from a bulk crystal, sourced from Manchester Nanomaterials, using the scotch tape method, and subsequently transferred, in three separate processes, onto a Si/SiO$_2$, a rough Si/SiO$_2$/Au, and a smooth Si/SiO$_2$/Au substrate. The rough Si/SiO$_2$/Au substrate was fabricated by the spattering of Au on a Si/SiO$_2$ substrate in a vacuum physical vapor deposition growth chamber (pressure $\sim$ 10$^{-3}$ Torr), while the smooth Si/SiO$_2$/Au substrate was prepared by standard thermal evaporation of Au. All PL and Raman measurements were performed in a back-scattering geometry using a micro-Raman spectrometer (Horiba LabRAM HR, Jovin Yvon), equipped with 600 lines/mm grating and a Peltier-cooled CCD detector. Excitation was provided by an air-cooled continuous wave Argon-ion laser with a wavelength of 488 nm (2.54 eV), focused onto the sample using a 50X objective lens with a numerical aperture of 0.75. The excitation power was $\sim$ 0.1 mW unless otherwise stated. A linear polarizer was used into the incident optics and a quarter-wave plate was added to both the incident and collection paths followed by a half wave plate along with a linear polarizer at the collection optics for circularly polarization resolved PL studies. Low-temperature PL spectra were recorded using a Linkam THMS600 temperature-controlled stage.

\begin{center}
\textbf{Section II : Room temperature PL spectra}
\end{center}

\begin{figure}[h]
\centering
  \includegraphics[height=4.5 cm]{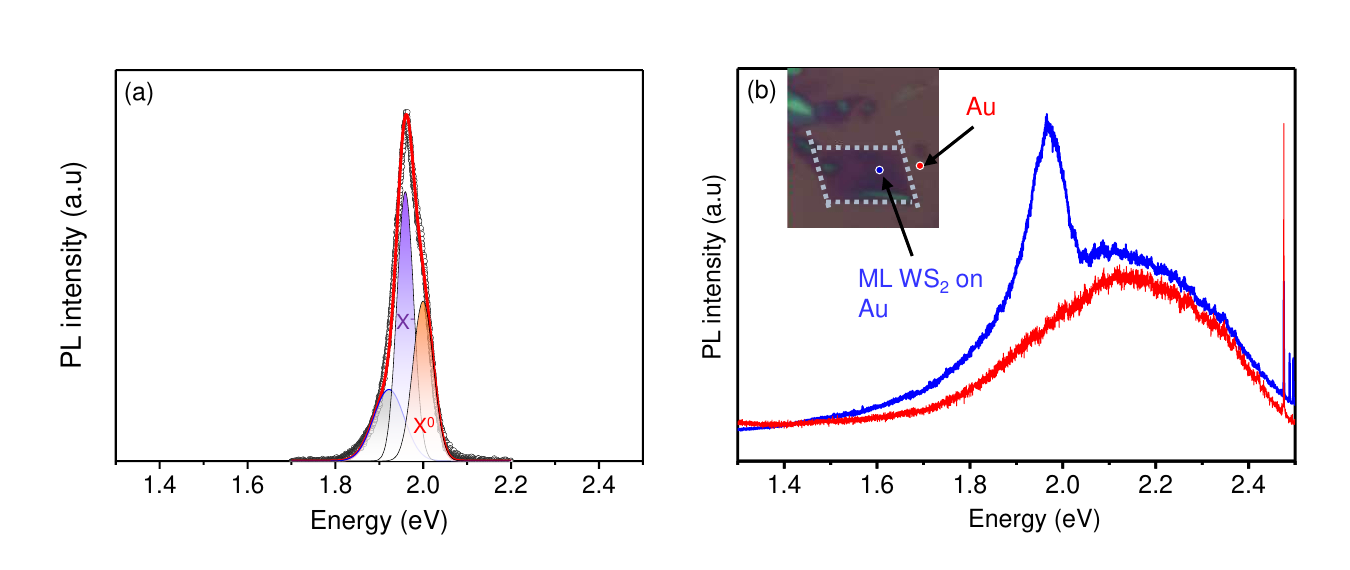}
  \caption{(a) Room-temperature PL spectrum of ML WS$_2$ on a Si/SiO$_2$ substrate. (b) Room-temperature PL spectra from ML WS$_2$ on the disordered Au film and from the sole disordered Au film, as indicated in the figure.}
\end{figure}

\begin{center}
\textbf{Section III : LSP-induced hot electron injection from Au substrate to ML WS$_2$}
\vspace{0.1 in}
\end{center}

\begin{figure}[h]
\centering
  \includegraphics[height=4.2 cm]{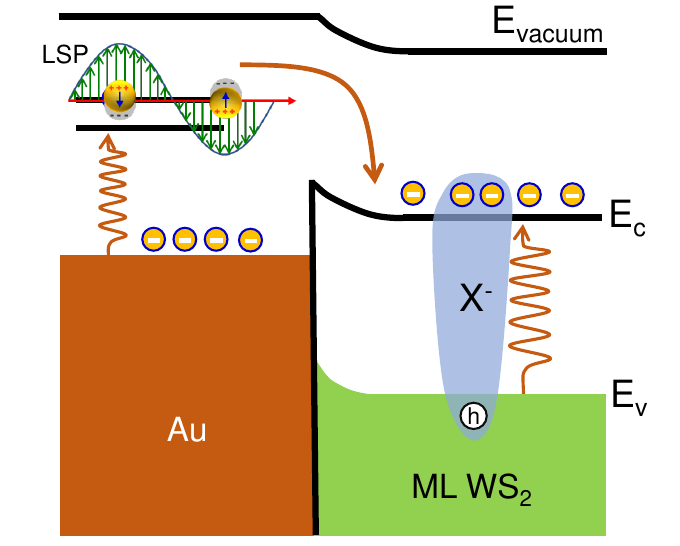}
  \caption{Schematic illustration of plasmon-assisted electron transfer from a rough Au substrate to ML WS$_2$. Optical excitation of localized surface plasmons (LSPs) at the Au surface generates energetic ("hot") electrons. These electrons are transferred across the Au/WS$_2$ interface into the conduction band (E$_c$) of ML WS$_2$, resulting in electron accumulation and n-type doping of the monolayer. The transferred electrons modify the carrier density and excitonic population in WS$_2$, favoring the formation of trions. E$_c$ and E$_v$ denote the conduction and valence band edges of WS$_2$, respectively.}
\end{figure}

\newpage
\begin{center}
\textbf{Section IV : Raman measurements}
\vspace{0.1 in}
\end{center}

\begin{figure}[h]
\centering
  \includegraphics[height=7cm]{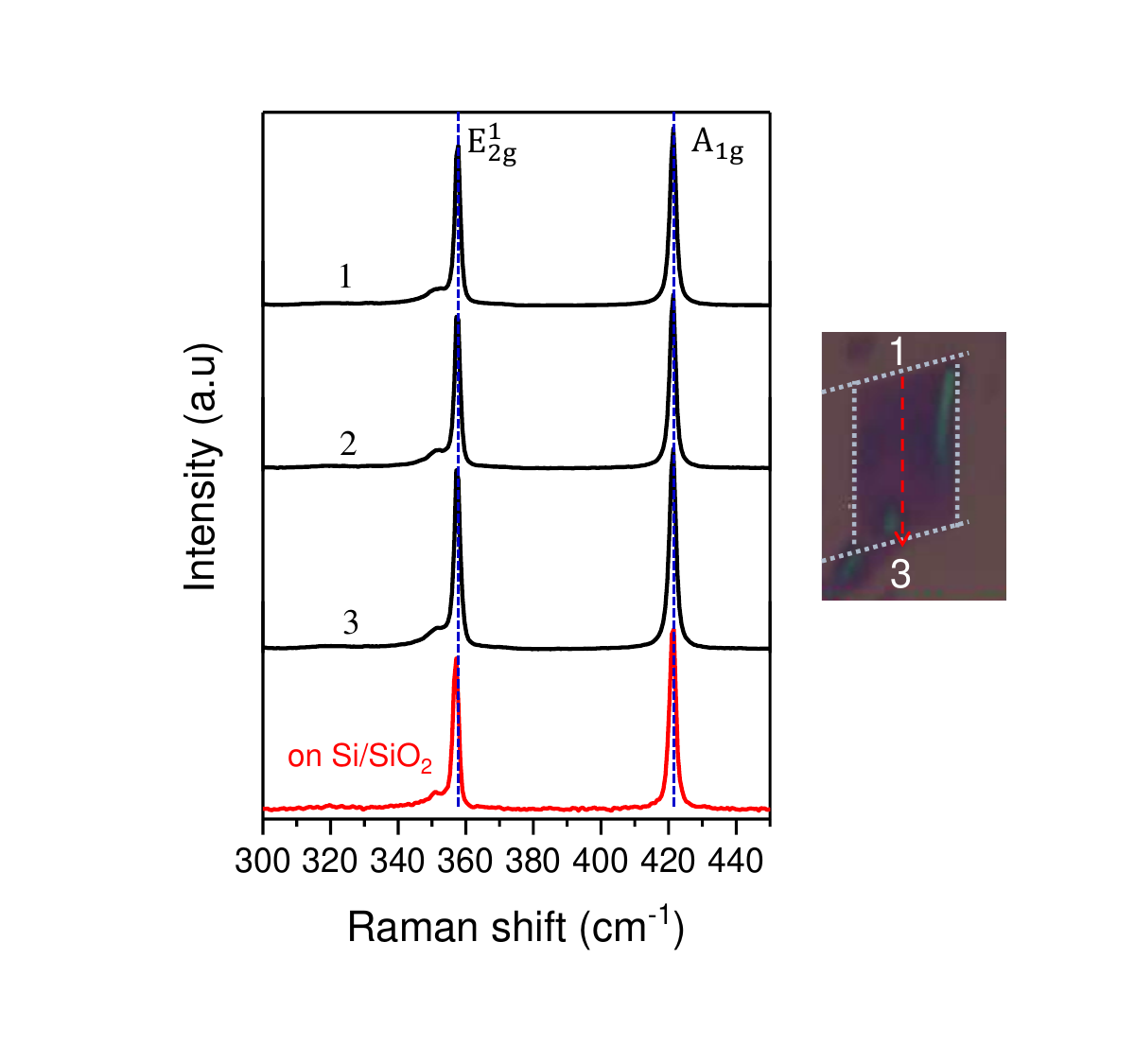}
  \caption{Raman spectra acquired at multiple positions along the line scan on the ML WS$_2$ supported on the rough Au film (black curves). The right panel indicates the locations of the measurement points on the Au-supported sample. For comparison, the Raman spectrum from an ML WS$_2$ sample on a Si/SiO$_2$ substrate is also shown (red curve).}
\end{figure}

To evaluate the presence of strain in the sample, we performed Raman measurements at multiple positions across the sample at room temperature using a 100× objective with a numerical aperture (NA) of 0.9. Under these conditions, the laser spot size is approximately 660 nm, which is smaller than that used in the photoluminescence (PL) measurements (50× objective with NA = 0.75). Consequently, the Raman measurements provide higher spatial resolution than the PL measurements and are therefore more sensitive to local strain variations. The Raman spectra consistently show the characteristic E$^1_{2g}$ and A$_{1g}$ modes at 357.5 cm$^{-1}$ and 421.4 cm$^{-1}$, respectively. Importantly, when Raman spectra were collected along a line scan spanning positions 1–3 (see Figure S3) across the sample, no measurable shift or broadening was observed in either Raman mode. Since the E$^1_{2g}$ mode is well known to be highly strain-sensitive, the absence of any detectable modulation in its peak position places a strict upper bound on strain in our sample \cite{wang2015strain,wang2020strain,blundo2020evidence,michail2023tuning}. Previous studies have shown that the E$^1_{2g}$ mode softens continuously under strain \cite{wang2015strain,wang2020strain,blundo2020evidence,michail2023tuning}. In contrast, our Raman data show neither softening nor broadening, indicating that the strain present in our sample is insignificant, which can affect exciton landscape. 

\begin{center}
\textbf{Section V : Topography}
\vspace{0.1 in}
\end{center}

We examined the local topography using AFM. The line-profile analysis in several regions (see Fig S4) of the sample reveals very small grain height variation. Such small topographic fluctuations are not expected to induce any appreciable strain in ML WS$_2$ \cite{blundo2020evidence}. 
 
\begin{figure}[h]
\centering
  \includegraphics[height=4.2cm]{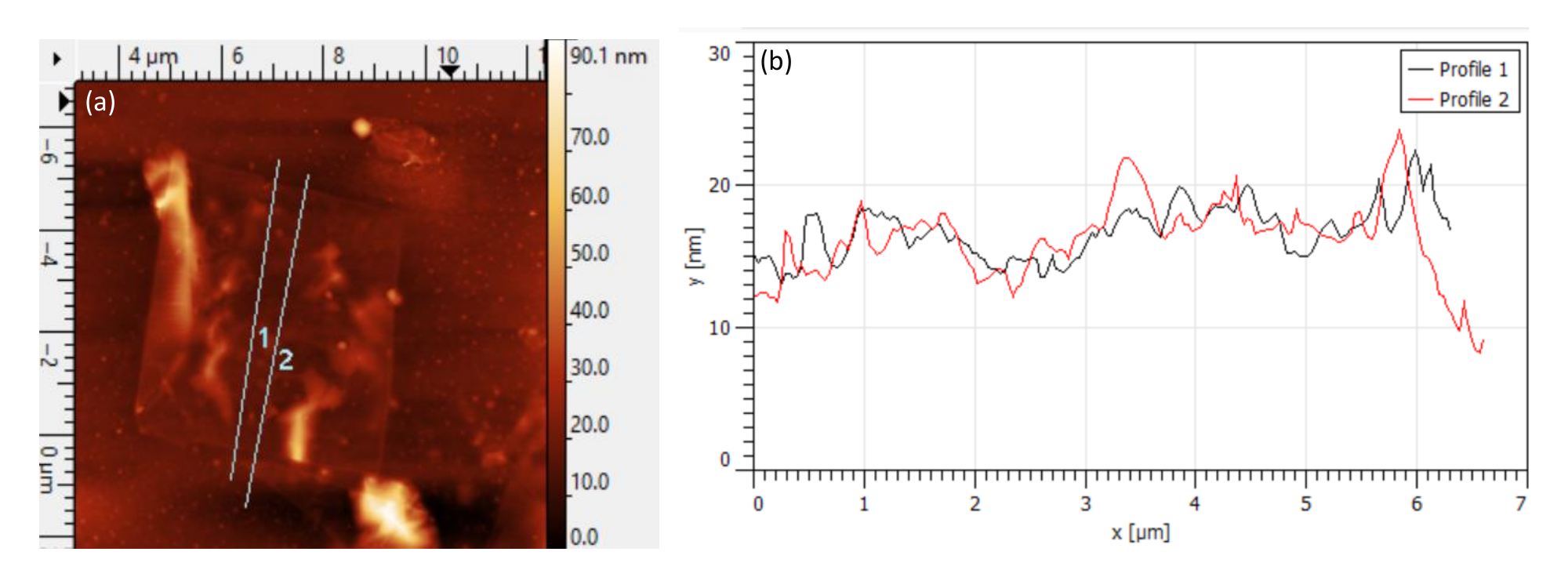}
  \caption{(a) shows AFM image with different linescans (1,2), which are plotted in the right panel (b).}
\end{figure}

\begin{center}
\textbf{Section VI : Laser power-dependent PL data}
\vspace{0.1 in}
\end{center}

\begin{figure}[h]
\centering
  \includegraphics[height=7 cm]{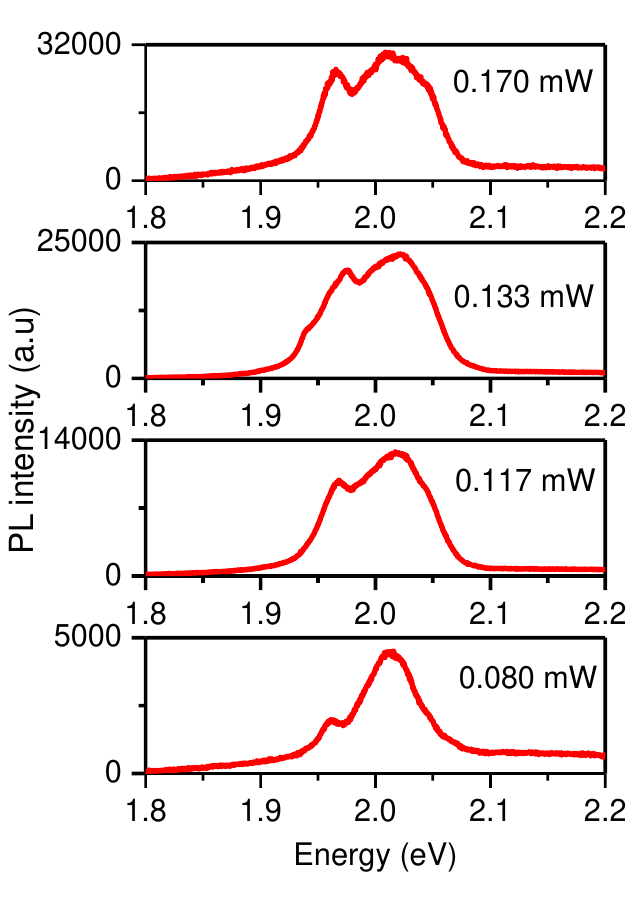}
  \caption{Laser power dependent PL spectra from ML WS$_2$ on a disordered Si/SiO$_2$/Au film at 83 K.}
\end{figure}

\begin{center}
\textbf{Section VII : Electrodynamics Simulation}
\vspace{0.1 in}
\end{center}

To examine how the measured surface morphology of the Au film influences the spatial distribution of optical fields, we performed finite-element simulations using the experimentally obtained AFM height map as the input geometry. The AFM data were interpolated onto a two-dimensional computational mesh, and a frequency-domain wave equation was solved under plane-wave illumination using a scattered-field formulation.  

Absorbing boundary conditions were applied at the outer edges of the simulation domain to minimize artificial reflections. The resulting field distributions were used to generate spatial maps of the out-of-plane near-field intensity above the surface, normalized to the incident field amplitude. These simulations are included to visualize morphology-dependent field localization and to support the discussion of enhanced radiative visibility of semi-dark trion on rough Au substrate.

 
\begin{figure}[h]
\centering
  \includegraphics[height=6.5cm]{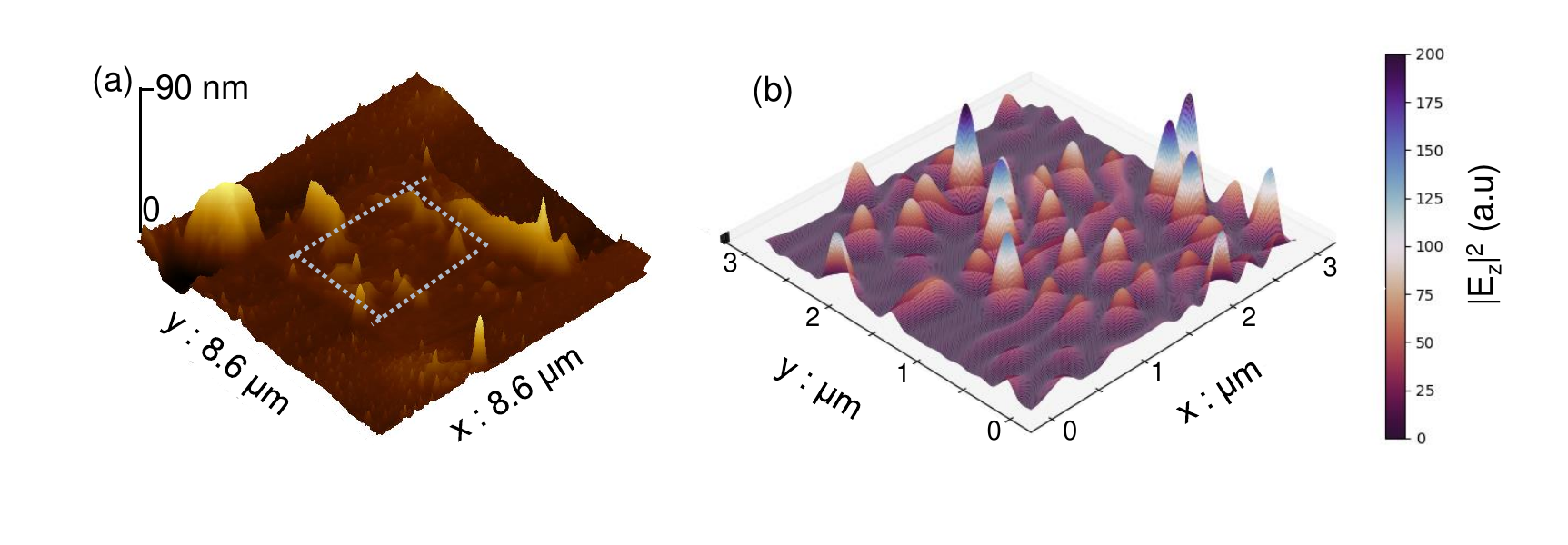}
  \caption{(a) AFM image of the sample, where the dotted lines indicate central part of the sample, which has been taken for the simulation, result of which is shown in (b).}
\end{figure}






\begin{center}
\textbf{Section VIII: Plasmon-assisted mechanism underlying the observed PL doublet}
\vspace{0.1 in}
\end{center}

The PL emission from a gold film exhibits a broad band spanning over a range of energies, contrasting with the distinct peaks exhibited by semiconductors. The functional advantage of the roughness of a gold film is in the enhancement of its PL emission intensity. The topography of the rough surface, characterized by bumps, islands and gaps, creates microscopic “hot spots” of intense localized electromagnetic fields via Localized Surface Plasmon Resonance, increasing thereby the PL absorption and radiative emission rates. In the case of a flat gold film, the PL emission is known to be quenched due to the dominance of nonradiative recombination pathways \cite{boyd1986photoinduced,ngoc2015plasmon,tan2016rough}. Fig. S7 compares our experimental plots of the PL emission intensity from rough and flat gold films on a Si/SiO$_2$ substrate. In the energy range of interest (below 2.2 eV) in which the PL doublet is formed, one finds that of the two cases, smooth and rough gold films, the emission intensity is higher in the latter case, indicating the prominent presence of localized surface plasmons.

\begin{figure}[h]
\centering
  \includegraphics[height=6 cm]{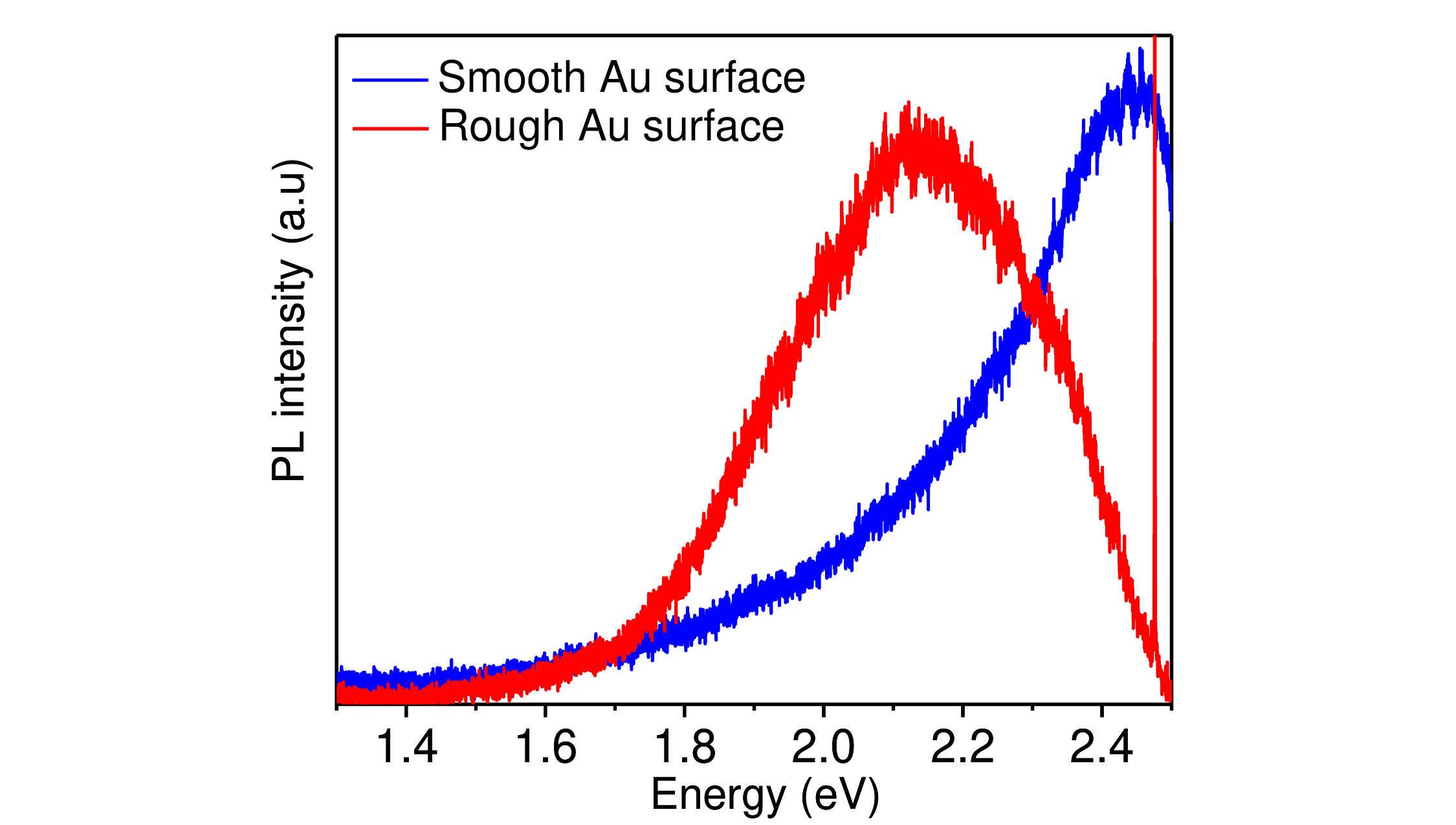}
  \caption{PL spectra from smooth and rough gold surface, as indicated in the figure.}
\end{figure}

A physical understanding of the doublet formation is based on a combination of
intervalley e-e scattering involving trions and localized surface plasmons generated in a rough gold film. This is borne out by the three plots shown in Fig. S8 in which one observes the PL doublet only when the gold film is rough.

\begin{figure}[h]
\centering
  \includegraphics[height=7.5 cm]{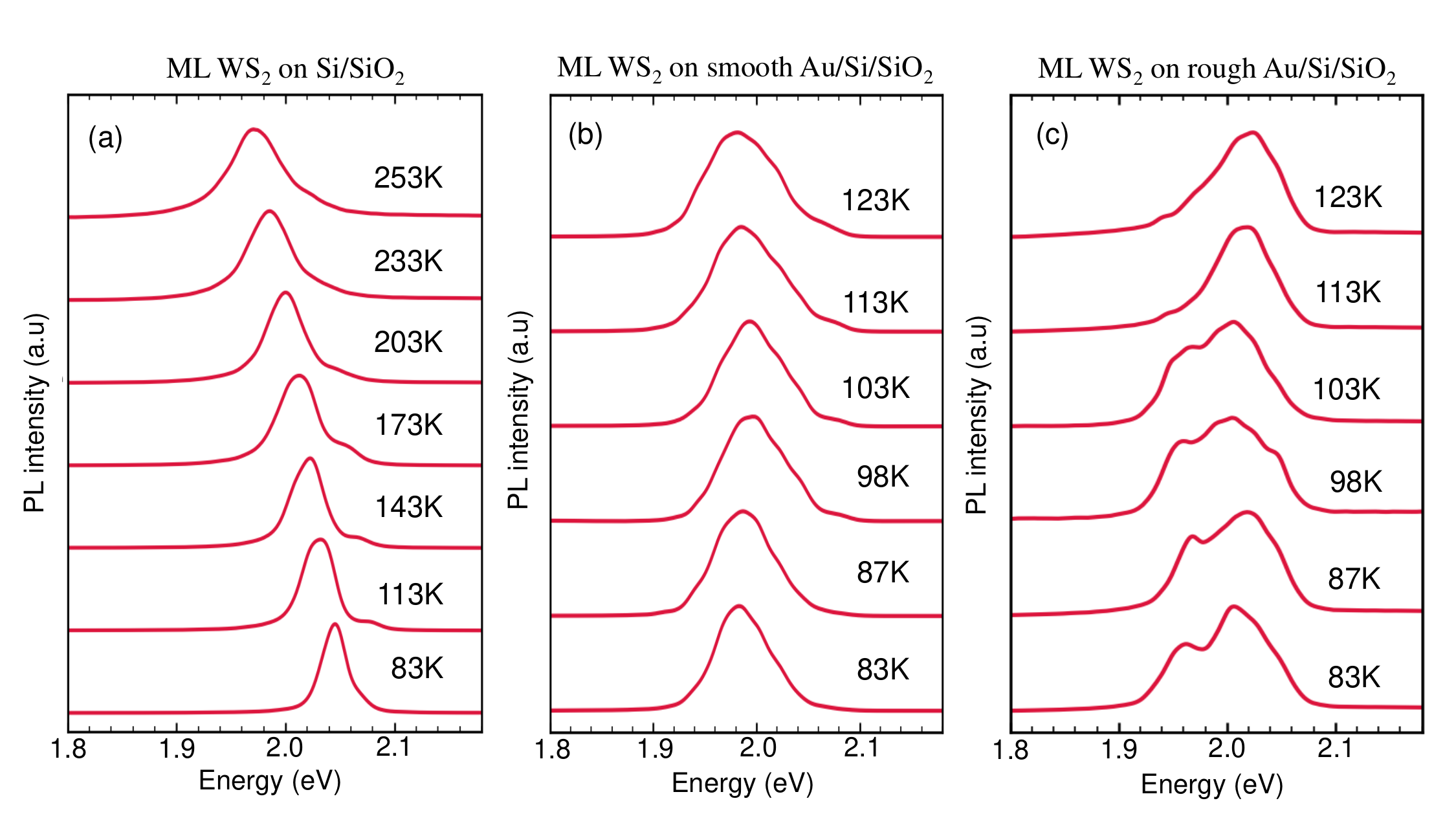}
  \caption{Temperature dependent PL spectra from ML WS$_2$ on Si/SiO$_2$ substrate (a), on smooth Au/Si/SiO$_2$ substrate (b), and on rough Au/Si/SiO$_2$ substrate (c).}
\end{figure}

\newpage
\begin{center}
\textbf{Section IX : Temperature-dependent PL spectra from ML WS$_2$ on disordered Si/SiO$_2$/Au substrate}
\vspace{0.1 in}
\end{center}

\begin{figure}[h]
\centering
  \includegraphics[height=9 cm]{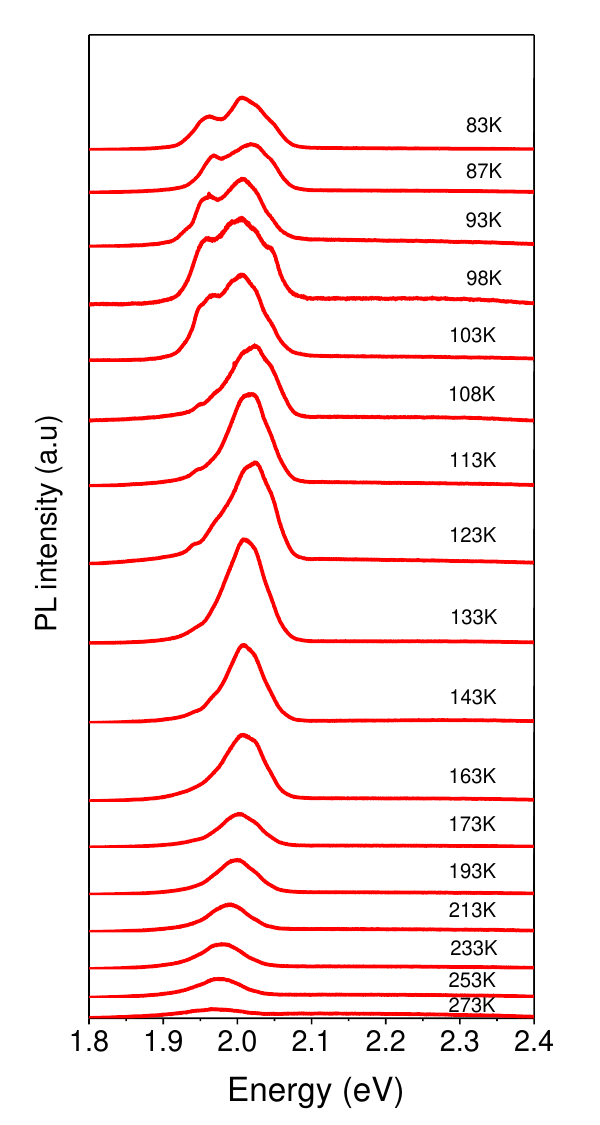}
  \caption{Temperature-dependent PL spectra from ML WS$_2$ on a disordered Si/SiO$_2$/Au film.}
\end{figure}

\begin{center}
\textbf{Section X : Integrated PL intensity as a function of temperature from ML WS$_{2}$ on Si/SiO$_{2}$ substrate}
\vspace{0.1 in}
\end{center}

\begin{figure}[h]
\centering
  \includegraphics[height=6cm]{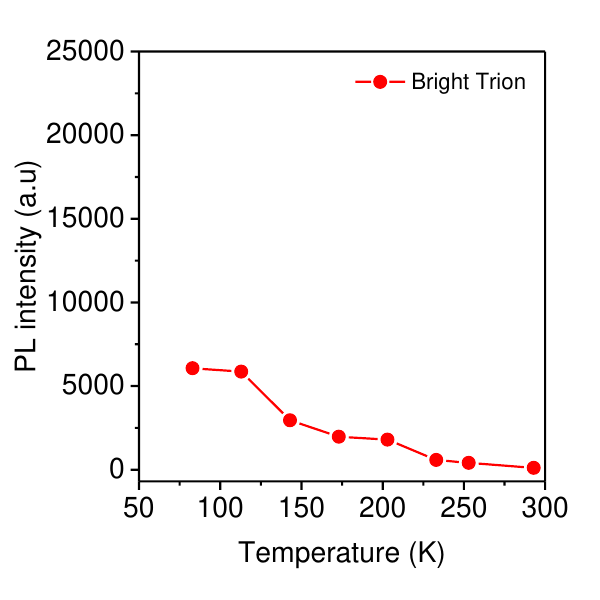}
  \caption{Temperature-dependent integrated PL intensity of bright trion from ML WS$_2$ on a Si/SiO$_2$ substrate.}
\end{figure}


\begin{center}
\textbf{Section XI : Fine structure of PL spectrum from ML WS$_2$ on Si/SiO$_2$/Au substrate at 83 K}
\vspace{0.1 in}
\end{center}

\begin{figure}[h]
\centering
  \includegraphics[height=7cm]{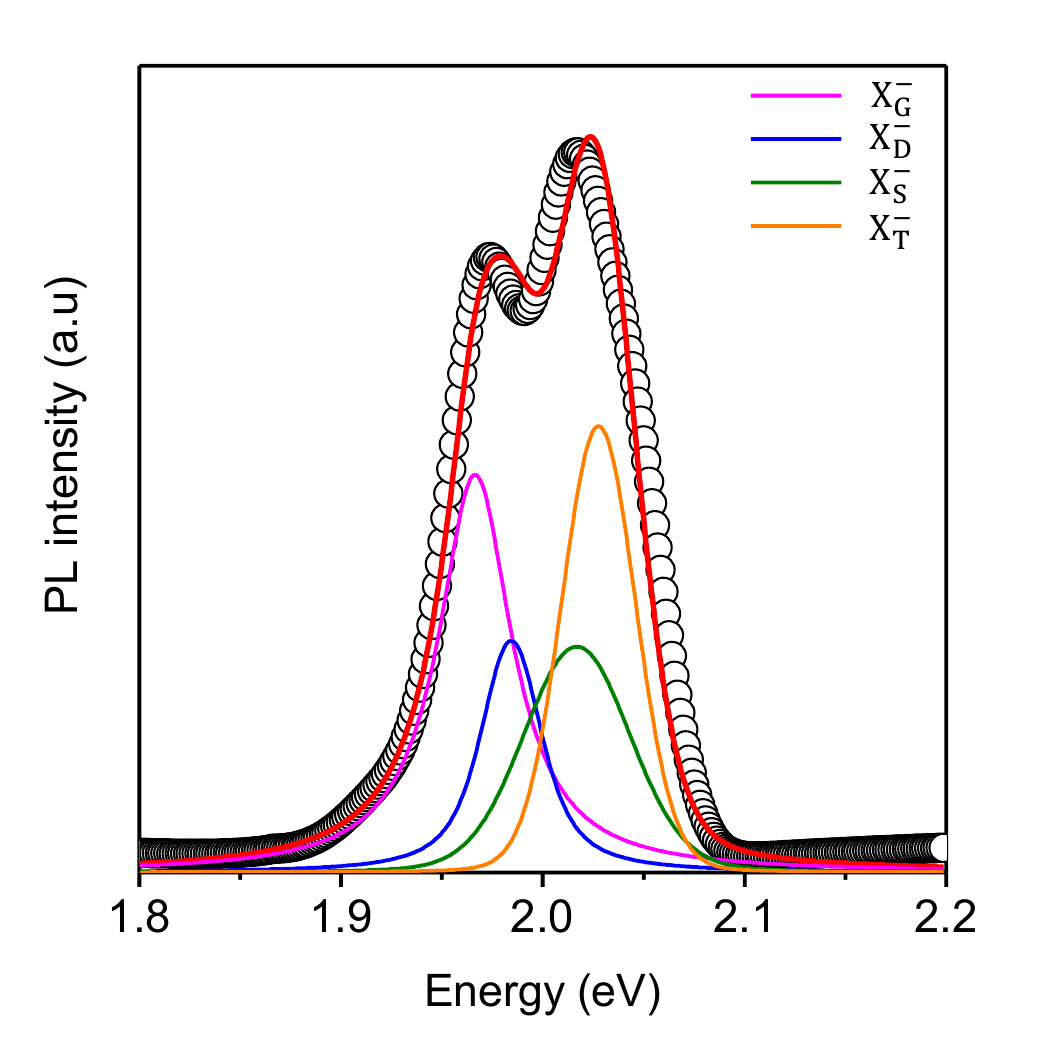}
  \caption{Deconvoluted PL spectrum of monolayer WS$_2$ on a disordered Si/SiO$_2$/Au substrate at 83 K, resolved into four components corresponding to the semi-dark gray trion (X$_{\mathrm{G}}^{-}$), the dark trion (X$_{\mathrm{D}}^{-}$), the spin-singlet bright trion (X$_{\mathrm{S}}^{-}$), and the spin-triplet bright trion (X$_{\mathrm{T}}^{-}$). The fitted curve (red line) closely matches the experimental data set (open circles).}
\end{figure}

A four-component analysis of the doublet at 83 K (Fig. S11) shows
overlapped PL signals from all the four trions with successive peak positions at 1.9691 eV, 1.9831 eV, 2.0148 eV and 2.0278 eV. These energies are red-shifted from the corresponding peak positions at 1.996 eV, 2,0078 eV, 2,028 eV and 2.034 eV, as reported in \cite{chand2023interaction} at 7 K. The peak energy difference between the dark and semi-dark trions, in our analysis of the PL signal, is obtained as 14
meV, close to the estimates reported earlier \cite{PhysRevB.111.155409,doi:10.1021/acs.nanolett.0c05021,chand2023interaction,tu2019experimental}. The splitting energy between the triplet and singlet trions is approximately 13 meV. Based on a similar component analysis, a value of 11 meV has been obtained previously \cite{Plechinger2016}.

The fitting was performed using constrained multi-peak fitting, in which the component positions, widths, and amplitudes were not allowed to vary arbitrarily, but were restricted to physically reasonable ranges guided by the expected fine-structure hierarchy. Thus, the four-peak representation was used as a model-assisted decomposition of the observed spectral envelope, rather than as direct proof that four isolated peaks are experimentally resolved.



\begin{center}
\textbf{Section XII : Circular polarization resolved PL spectra from ML WS$_{2}$ on Si/SiO$_{2}$ substrate at 83 K}
\vspace{0.1 in}
\end{center}

\begin{figure}[h]
\centering
  \includegraphics[height=8 cm]{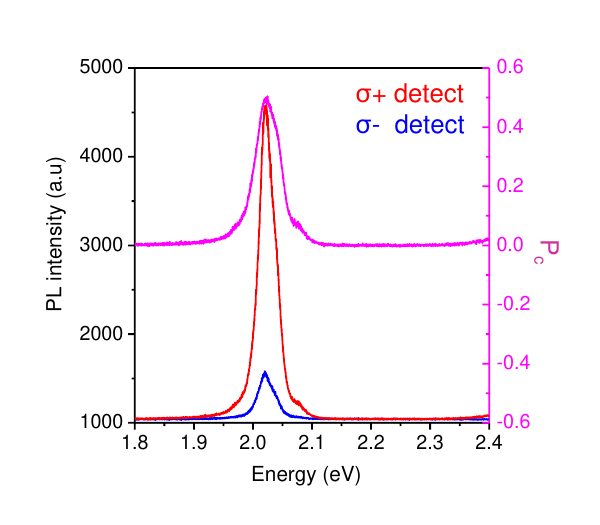}
  \caption{Circularly polarization resolved PL spectra from a ML WS$_{2}$ on Si/SiO$_{2}$ substrate under $\sigma+$ circularly polarized excitation at 488 nm. Emission is analyzed in both $\sigma+$ and $\sigma-$ detection channels. The degree of circular polarization ($P_c$) is plotted as a function of emission energy. A $P_c$ of approximately 50 \% is observed for the bright trion emission and around 7 \% for the bright neutral exciton.}
\end{figure}


\bibliography{sample}

\section*{Acknowledgments}
I.B. acknowledges the support of NASI, Allahabad, India, under the Honorary Scientist Scheme. A.S. thanks the Anusandhan National Research Foundation (ANRF), India, for their financial support (File No. ANRF/ARG/2025/010\\454/PS). The authors thank Subhajit Mahapatra for some additional measurements. The authors acknowledge the assistance of Suvadip Masanta, Chumki Nayak, Pritam Sinha, and Prithwiraj Majhi. The authors thank the Central Instrumental Facility at S. N. Bose National Centre for Basic Sciences (SNBNCBS), Kolkata, for assisting with the AFM measurements. The authors thank Prof. Xavier Marie for useful discussions.

\end{document}